\documentclass[12pt]{iopart}
\usepackage{amsfonts}
\begin{document}
\title{Can Mathisson-Papapetrou equations give clue to some problems in astrophysics?}
\author{Roman  Plyatsko}
\address{ Pidstryhach Institute of Applied Problems in Mechanics and
Mathematics,\\ Ukrainian National Academy of Sciences, 3-b Naukova
Str.,\\ Lviv, 79060, Ukraine}

\ead{plyatsko@lms.lviv.ua}

\begin{abstract}

First, we stress that for correct description of highly
relativistic fermions in a gravitational field it is necessary to
have an equation which in the limiting transition to the classical
(non-quantum) case corresponds to the exact Mathisson-Papapetrou
equations. According to these equations the spin in general
relativity is Fermi-transported, and the parallel transport of
spin is realized only in some approximation. The traditional
general-relativistic Dirac equation (1929) is based on the
parallel transported spinors and does not ensure the correspondent
transition. Second, because in the range of very high velocity
(close to the speed of light) of a spinning particle relative to
the Schwarzschild or Kerr sources the Mathisson-Papapetrou
equations have the solutions which reveal that the spin-gravity
interaction acts as a strong antigravity force, we suppose that
this fact can be useful for explanation some astrophysical
phenomena. Some association with the OPERA results is possible.

\end{abstract}

%\submitto{\CQG}
% \pacs{ 04.20.-q, 95.30.Sf}

%\maketitle
%\section {Introduction}

%\section {General form of the MP equations at Mathisson-Pirani and
%Tulczyjew-Dixon conditions}

The Mathisson-Papapetrou equations are known from 1937 as the
equations which describe motions of a spinning test particle
(rotating test body) in a gravitational field in the framework of
general relativity [1]. These equations can be written as
\begin{equation}\label{1}
\frac D {ds} \left(mu^\lambda + u_\mu\frac {DS^{\lambda\mu}}
{ds}\right)= -\frac {1} {2} u^\pi S^{\rho\sigma}
R^{\lambda}_{~\pi\rho\sigma},
\end{equation}
\begin{equation}\label{2}
\frac {DS^{\mu\nu}} {ds} + u^\mu u_\sigma \frac {DS^{\nu\sigma}}
{ds} - u^\nu u_\sigma \frac {DS^{\mu\sigma}} {ds} = 0,
\end{equation}
where $u^\lambda\equiv dx^\lambda/ds$ is the particle's
4-velocity, $S^{\mu\nu}$ is the tensor of spin, $m$ and $D/ds$
are, respectively, the mass and the covariant derivative with
respect to the particle's proper time $s$;
$R^{\lambda}_{~\pi\rho\sigma}$ is the Riemann curvature tensor
(units $c=G=1$ are used); here and in the following, latin indices
run 1, 2, 3 and greek indices 1, 2, 3, 4; the signature of the
metric (--,--,--,+) is chosen.

While investigating the solutions of equations (1), (2), it is
necessary to add a supplementary condition  in order to choose an
appropriate trajectory of the particle's center of mass. Most
often conditions
\begin{equation}\label{3}
S^{\lambda\nu} u_\nu = 0
\end{equation}
or
\begin{equation}\label{4}
S^{\lambda\nu} P_\nu = 0
\end{equation}
are used, where
\begin{equation}\label{5}
P^\nu = mu^\nu + u_\lambda\frac {DS^{\nu\lambda}}{ds}
\end{equation}
is the 4-momentum. The condition for a spinning test particle
\begin{equation}\label{6} \frac{|S_0|}{mr}\equiv\varepsilon\ll 1
\end{equation}
must be taken into account as well, where  $|S_0|=const$ is the
absolute value of spin, $r$ is the characteristic length scale of
the background space-time (in particular, for the Kerr metric $r$
is the radial coordinate), and $S_0$ is determined by the
relationship
\begin{equation}\label{7} S_0^2=\frac12
S_{\mu\nu}S^{\mu\nu}.
\end{equation}
Instead of exact MPD equations (1) their linear spin approximation
\begin{equation}\label{8}
m\frac D {ds} u^\lambda = -\frac {1} {2} u^\pi S^{\rho\sigma}
R^{\lambda}_{\,\,\,\pi\rho\sigma}
\end{equation}
is often considered. The long list of publications devoted to the
Mathisson-Papapetrou equations is presented, for example, in
[2--4].

Our purpose is to draw attention to the two points concerning the
strict Mathisson-Papapetrou equations. The first is connected with
the general-relativistic Dirac equation, which was obtained in
1929 [5], i.e., eight years before the Mathisson-Papapetrou
equations. Later it was shown in many papers that the
Mathisson-Papapetrou equations are, in certain sense, the
classical approximation of the Dirac equation [6]. The main step
in obtaining the general-relativistic Dirac equation in the curved
spacetime consists in introduction the notion of the parallel
transport for spinors as a generalization of this notion for
tensors. However, if one want to satisfy the principle of
correspondence between the general-relativistic Dirac equation and
the Mathisson-Papapetrou equations, it is necessary to take into
account the known fact that according to the Mathisson-Papapetrou
equations the spin of a test particle is transported by Fermi, nor
parallel transported (we underline that this fact was unknown in
19290. The Fermi transport coincides with the parallel transport
only in some approximation, when a world line of a spinning
particle practically coincides with the corresponding geodesic
line, for example, in the post-Newtonian approximation. In
general, by the principle of correspondence, it is necessary to
know how to write the Dirac equation in the curved spacetime with
the Fermi transport in the limiting transition to the classical
(nonquantum) description. One can suppose that for this aim it is
sufficiently to introduce the Fermi transport for spinors, instead
of their parallel transport (in this sense an attempt was
discussed in [7]). However, in common sense, the notation
"Fermi-transported spinor" cannot be introduced without violation
of the Lorentz invariance. By the way, in this context it is
interesting that for last years the possibility of the Lorentz
invariance violation  is discussed in the literature from
different points of view.

In any case (with violation of the Lorentz invariance or not), it
is necessary to propose a more exact equation for fermions in
gravitational field than the usual general-relativistic Dirac
equation (probably, this equation must be nonlinear in the $\psi$-
function [7]).

The second point of importance connected with the
Mathisson-Papapetrou equations is the properties of their
solutions which describe highly relativistic motions of a spinning
particle in the Schwarzschild or Kerr backgrounds. Namely, if the
velocity of a particle relative to the Schwarzschild or Kerr mass
corresponds to the relativistic Lorentz $\gamma$ -- factor of
order $1/\sqrt{\varepsilon}$, where $\varepsilon$ is determined by
(6), the trajectory of this particle can significantly differ from
the corresponding trajectory of a spinless particle, i.e, from the
geodesic line [3, 4] (in these papers same cases are calculated
when $r$ is not much grater than the horizon radius). That is, the
Mathisson-Papapetrou equations predict an interesting phenomenon:
for highly relativistic spinning particles gravity becomes
antigravity, at some correlation of signs of the spin and the
particle's orbital velocity (for another correlation of these
signs the spin-gravity interaction acts as an strong attractive
force).

The necessary value of the Lorentz factor is less for the
particles with higher ratio spin/mass, if $r$ in (6) is fixed. For
example, this value is less for neutrino than for electron or
proton (some numerical estimates are presented in [3, 8, 9]). By
the way, for neutrino near the Earth surface the value 
$1/\sqrt{\varepsilon}$ is of order $3\times 10^6$. That is, the
Lorentz factor for neutrinos in the OPERA experiments is much
higher.

\end{document}